\documentclass[11pt,journal,onecolumn]{IEEEtran}
\usepackage{epsfig,graphicx,psfrag}
\usepackage{amsfonts,amssymb,amsmath}

\newtheorem{lemma}{Lemma}
\newtheorem{theorem}{Theorem}
\newtheorem{definition}{Definition}

\newcommand{\beq}{\begin{eqnarray}}
\newcommand{\eeq}{\end{eqnarray}}
\newcommand{\tends}{{\to}}
\newcommand{\cq}{{\cal{Q}}}

\begin{document}
\title{Throughput and Delay in Random Wireless Networks with Restricted 
Mobility}
\author{James Mammen and Devavrat Shah
\thanks{J. Mammen is with the department of Electrical Engg. at Stanford 
University. D. Shah is with the departments of EECS and ESD at
MIT. Emails: jmammen@stanford.edu; devavrat@mit.edu.}}
\maketitle

\begin{abstract} 
Grossglauser and Tse (2001) introduced a mobile random network model
where each node moves independently on a unit disk according to a
stationary uniform distribution and showed that a throughput of
$\Theta(1)$ is achievable.  El Gamal, Mammen, Prabhakar and Shah
(2004) showed that the delay associated with this throughput scales as
$\Theta\left(n\log n\right)$, when each node moves according to an
independent random walk.  In a later work, Diggavi, Grossglauser and
Tse (2002) considered a random network on a sphere with a restricted
mobility model, where each node moves along a randomly chosen great
circle on the unit sphere. They showed that even with this
one-dimensional restriction on mobility, constant throughput scaling
is achievable. Thus, this particular mobility restriction does not
affect the throughput scaling. This raises the question whether this
mobility restriction affects the delay scaling.

This paper studies the delay scaling at $\Theta(1)$ throughput for a
random network with restricted mobility. First, a variant of the
scheme presented by Diggavi, Grossglauser and Tse (2002) is presented
and it is shown to achieve $\Theta(1)$ throughput using different (and
perhaps simpler) techniques. The exact order of delay scaling for this
scheme is determined, somewhat surprisingly, to be of $\Theta(n\log
n)$, which is the same as that without the mobility restriction. Thus,
this particular mobility restriction \emph{does not} affect either the
maximal throughput scaling or the corresponding delay scaling of the
network. This happens because under this 1-D restriction, each node is
in the proximity of every other node in essentially the same manner as
without this restriction.
\end{abstract}


\section{Introduction}
\label{s:intro}
Gupta and Kumar \cite{GK} introduced a random network model for
studying throughput scaling in a fixed wireless network (that is, when
the nodes do not move). They defined a random network to consist of
$n$ nodes where each node is distributed uniformly and independently
on the unit sphere in ${\mathbb R}^3$. The network has $n/2$ distinct
source-destination pairs formed at random. Each node can transmit at
$W$ bits-per-second provided that the interference is sufficiently
small. They showed that in such a random network the throughput scales
as $\Theta(1/\sqrt{n\log n})$ per source-destination (S-D) pair.

Grossglauser and Tse \cite{GT} showed that by allowing the nodes to
move, the throughput scaling changes dramatically.  Indeed, if node
motion is independent across nodes and has a uniform stationary
distribution, a constant throughput scaling ($\Theta(1)$) per S-D pair
is feasible. This raised the question: what kind of mobility is
necessary for achieving constant throughput scaling? Diggavi,
Grossglauser and Tse \cite{DGT} considered a restricted mobility model
where each node is allowed to move along a randomly chosen great
circle on the unit sphere with a uniform stationary distribution along
the great circle. They showed that a constant throughput per S-D pair
is feasible even with this restricted mobility model. Thus they
established that node motion with a stationary distribution on the
entire network area is not necessary for achieving constant throughput
scaling.

El Gamal, Mammen, Prabhakar and Shah \cite{EMPS} (see
\cite{EMPS_TDto_fluid} for complete details) determined the
throughput-delay trade-off for both fixed and mobile wireless
networks.  In particular, it was shown that for mobile networks at
throughput of $\Theta(1)$, the delay is $\Theta(n\log n)$.  For mobile
networks, the mobility model consisted of each node moving
independently according to a symmetric random walk on a
$\sqrt{n}\times \sqrt{n}$ grid on the unit torus.

The constant throughput scaling result of \cite{DGT} for a network
with restricted mobility raises the question whether the high
throughput in spite of restricted mobility is at the expense of
increased delay. Motivated by this question, we study the delay
scaling for constant throughput scaling in a network with restricted
mobility. Somewhat surprisingly, we find that delay scaling is not
affected by this mobility restriction either. That is, delay scales as
$\Theta(n\log n)$, which is the same as the delay scaling when
mobility is not restricted. This paper is a consolidation of the
preliminary work presented in \cite{MS04}.

This seemingly surprising result can be explained as follows. Since
there are $n$ nodes in a network of constant area, the neighborhood of
each node is $\Theta(1/n)$. Based on this, let us say that two nodes
{\em meet} or are {\em neighbors} when they are within a distance of
$\Theta\left(1/\sqrt{n}\right)$. The following condition ensures
constant throughput scaling in the mobile network models presented in
\cite{GT}, \cite{DGT} and this paper: {\em for $\Theta(1/n)$ fraction of the 
time, each node is a neighbor of every other node with only 
$\Theta(1)$ other nodes in its neighborhood}. This ensures that the total network
throughput is $\Theta(n)$ and that it is distributed evenly among the
$n/2$ S-D pairs, so that the throughput is $\Theta(1)$. Delay is
determined by the first and second moments of the inter-meeting time
of the nodes. In the case of unrestricted mobility, the inter-meeting
time of any two nodes is equivalent to the inter-visit time to state
$(0,0)$ for a 2-D random walk on a $\sqrt{n}\times\sqrt{n}$ grid. In
the restricted mobility case also the inter-meeting time turns out to
be equivalent to the inter-visit time to state $(0,0)$ for a slightly
different random walk. However the first two moments are still of the
same order and hence the queueing delay is the same, leading to the
same delay scaling. As a result, even with this particular mobility
restriction, the maximal throughput scaling and the corresponding
delay scaling remain unchanged.

The rest of the paper is organized as follows. In
Section~\ref{s:model_defs}, we introduce the random mobile network
model, some definitions and notation. In
Section~\ref{s:scheme_Tput}, we present a scheme using random relaying
and show that it achieves constant throughput scaling. In
Section~\ref{s:delay}, we show that the delay for this scheme is
$\Theta(n\log n)$ using results which are proved in
Section~\ref{s:rem_proofs}. The proof of delay of $\Theta(n\log n)$
consists of analyzing a queue at a relay node in two parts. The first
part presented in Section~\ref{s:delay} identifies an i.i.d. component
that is embedded in the arrival and service processes of the
queue. The second part breaks the dependence between the arrival and
departure processes by introducing a virtual Bernoulli server. The
queueing analysis that follows is carried out in
Section~\ref{s:rem_proofs}.

\section{Models and Definitions}
\label{s:model_defs}
In this section, we present the network model, and the definitions of
the performance metrics -- throughput and delay. We begin by reminding
the reader of the order notation: (i) $f(n)=O(g(n))$ means that there
exists a constant $c$ and integer $N$ such that $f(n)
\leq cg(n)$ for $n > N$. (ii) $f(n)=o(g(n))$ means that $\lim_{n\to
\infty} f(n)/g(n) = 0$. (iii) $f(n)=\Omega(g(n))$ means that
$g(n)=O(f(n))$, (iv) $f(n)=\omega(g(n))$ means that
$g(n)=o(f(n))$. (v) $f(n)=\Theta(f(n))$ means that
$f(n)=O(g(n));~g(n)=O(f(n))$.

Now let us recall what is meant by the uniform distribution of great
circles on a sphere. Let $S^2$ denote the surface of a sphere in
${\mathbb R}^3$ with unit area. For $x\in S^2$, let $x'
\in S^2$ be the diametrically opposite point of $x$. Let $G(x)$ denote
the great circle obtained by the intersection of $S^2$ with the plane
passing through the center of $S^2$ and perpendicular to the line
$xx'$. Let $x$ be called the pole of $G(x)$. If the pole of a great
circle is chosen according to a uniform distribution on $S^2$ then the
great circle is said to have a uniform distribution.

\begin{definition}[Natural random walk] A natural random walk on a 
discrete torus of size $m$ is the process $S(t)\in\{0,\ldots,m-1\},
\:t=0,1,\ldots$, such that $S(0)$ is uniformly distributed over
$\{0,\ldots,m-1\}$ and $S(t+1)$ is equally likely to be any element of
$\{S(t), S(t) - 1 \mod m, S(t)+1 \mod m\}$.
\end{definition}

This differs from a simple random walk, where $S(t+1)$ is equally
likely to be any element of $\{S(t)-1\mod m, S(t)+1\mod m\}$. Since we
are interested only in scaling results, we use the terms (simple)
random walk and natural random walk interchangeably.

\begin{figure}[htpb]
\begin{center}
\begin{psfrags}
\psfrag{1}[l]{$i$}
\psfrag{2}[l]{$j$}
\psfrag{z12}[r]{$z_{ij}$}
\psfrag{C12}[r]{${\cal C}_{ij}$}
\includegraphics[width=3in,angle=0]{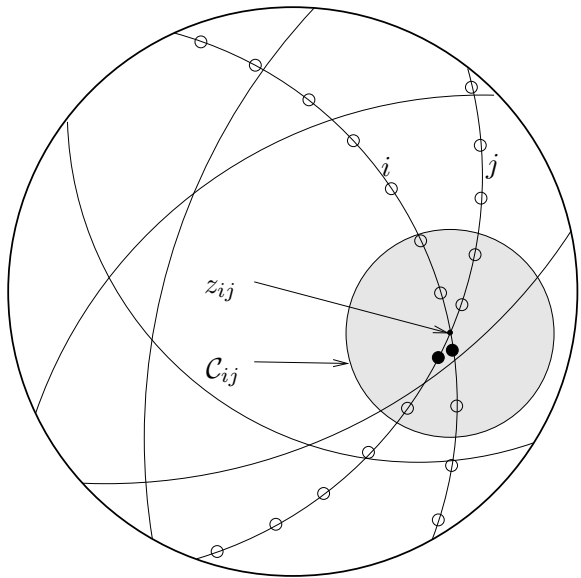}
\end{psfrags} 
\caption{A realization of the random network model. Only the 
lattice points on the great circles of nodes $i$ and $i$ are
shown. The intersection of their great circles is $z_{ij}$. The shaded
circle is ${\cal C}_{ij}$ and $i$ and $j$ become neighbors when they
are at the two dark lattice points.}
\label{f:1dmodel}
\end{center}
\end{figure}

\begin{definition}[Random network]
The random network consists of $n$ nodes that are split into $n/2$
distinct source-destination (S-D) pairs at random. Time is slotted for
transmission. Associated with each node is a great circle of $S^2$
chosen independently according to a uniform distribution.

The great circle of each node has $\sqrt{n}$ equidistant lattice
points numbered from $0$ to $\sqrt{n}-1$ placed on it arbitrarily
resulting in a one-dimensional discrete torus of size $\sqrt{n}$.
Each node moves according to a natural random walk on these lattice
points on its great circle.
Figure~\ref{f:1dmodel} shows a realization of the random network
model. Note that since the sphere has unit area, its radius is
$1/2\sqrt{\pi}$. Hence each great circle has perimeter $\sqrt{\pi}$
because of which the distance between two adjacent lattice points is
$\sqrt{\pi/n}$.
\end{definition}

Let the distance on the sphere between nodes $i$ and $j$ be denoted by
$d(i,j)$. We assume the Relaxed Protocol model \cite{EMPS} similar to
the Protocol model in \cite{GK} for successful transmission.
\begin{definition}[Relaxed Protocol Model]
A transmission from node $i$ to node $j$ is successful if for any
other simultaneously transmitting node $k$,
\[d(k,j) \ge (1+\Delta)d(i,j)\]
for some $\Delta>0$. If a transmission is successful then
communication occurs at a constant rate of $W$ bits-per-second. 
For simplicity, we assume that time-slots are of unit length so that
when a successful transmission occurs a packet of size $W$ is
communicated.
\end{definition}

In the other commonly used model (e.g., \cite{GT}, \cite{GK},
\cite{DGT}), known as the {\em Physical} model, a transmission is
successful if the Signal to Interference and Noise Ratio (SINR) is
greater than some constant. It is well known \cite{GK} that the {\em
Protocol} model is equivalent to the {\em Physical} model when each
transmitter uses the same power.

The differences between this model and the model in \cite{DGT} are:
(i) the Relaxed Protocol model is used instead of the Physical model,
and (ii) each node is assumed to move according to a natural random
walk instead of just a stationary, ergodic motion with uniform
stationary distribution on the great circle. However, this model has
the same 1-D mobility restriction. Further, the proofs clearly show
that the assumption of mobility according to a natural random walk is
not necessary for achieving constant throughput scaling and is used
only for computing delay.

\begin{definition}[Scheme]
A scheme $\Pi$ for a random network is a sequence of communication
policies, $\left(\Pi_n\right)$, where policy $\Pi_n$ determines how
communication occurs in a network of $n$ nodes.
\end{definition}

\begin{definition}[Throughput of a scheme]
Let $B_{\Pi_n}(i,t)$ be the number of bits of S-D pair $i, 1\leq i\leq
n/2$, transferred in $t$ time-slots under policy $\Pi_n$. Note that
this could be a random quantity for a given realization of the
network. Scheme $\Pi$ is said to have throughput $T_\Pi(n)$ if
$\exists$ a sequence of sets $A_\Pi(n)$ such that
\[ A_\Pi(n)=\left\{\omega: \min_{1\le i\le n/2} \lim\inf_{t\to\infty}
\frac{1}{t}B_{\Pi_n}(i,t) \ge T_\Pi(n)\right\} \]
and $P\left(A_\Pi(n)\right) \tends 1$ as $n \tends \infty$.
\end{definition}
We allow randomness in policies. Hence, $P\left(A_\Pi(n)\right)$
denotes the probability of $A_\Pi(n)$ over the joint probability space
that captures randomness in the policy as well as the random network
instance. We say that event $A$ occurs with high probability ({\em
whp}) if $P(A)\tends 1$ as $n\tends \infty$.

\begin{definition}[Delay of a scheme]
The delay of a packet is the time it takes for the packet to reach its
destination after it leaves the source. Let $D_{\Pi_n}^i(j)$ denote
the delay of packet $j$ of S-D pair $i$ under policy $\Pi_n$, then the
sample mean of delay for S-D pair $i$ under $\Pi_n$ is
\[ \bar{D}_{\Pi_n}^i = \limsup_{k\tends \infty}
\frac{1}{k} \sum_{j=1}^{k} D_{\Pi_n}^i(j).\]
The average delay over all S-D pairs for a particular realization of
the random network is then
\[\bar{D}_{\Pi_n} = \frac{2}{n} \sum_{i=1}^{n/2} \bar{D}_{\Pi_n}^i.\]
The delay for a scheme $\Pi$ is the expectation of the average delay
over all S-D pairs, i.e.,
\[ D_{\Pi}(n) = E[\bar{D}_{\Pi_n}] 
= \frac{2}{n} \sum_{i=1}^{n/2} E[\bar{D}_{\Pi_n}^i].\]
\end{definition}

Now observe that some realizations of the random network may result in
the configuration of nodes being such that it is not possible to
achieve constant throughput scaling. Hence we first define a typical
configuration which captures the fact that the distribution of great
circles is sufficiently uniform everywhere on the sphere. We need some
notation to introduce this definition.

Let $G_i$ denote the great circle of node $i \in \{1,\dots,n\}$.  For
any two nodes $i \neq j$, $G_i$ and $G_j$ are not identical with
probability $1$ under the random network model. Two distinct great
circles must intersect in exactly two points. For each pair $i\neq j$,
select one of the two distinct intersection points of $G_i$ and $G_j$
uniformly at random and call it $z_{ij}$. Let ${\mathcal C}_{ij}$
denote the disk on the sphere centered at $z_{ij}$ with radius
$(2+\Delta)\sqrt{\pi/n}$. See Figure~\ref{f:1dmodel} for an
illustration.

\begin{definition}[Typical configuration]
A configuration (i.e., realization of the random network) is said to
be {\em typical} if the number of great circles passing through each
${\cal C}_{ij}$ is $\Theta\left(\sqrt{n}\right)$.
\end{definition}

\begin{definition}[Neighbor]
We say that nodes $i$ and $j$ are {\em neighbors} at time $t$ if both
nodes $i$ and $j$ are at the lattice points of their respective great
circles that are closest to $z_{ij}$.
\end{definition}
In Figure~\ref{f:1dmodel}, the lattice points for nodes $i$ and $j$
that are closest to $z_{ij}$ have been darkened. Under the random walk
model, it is possible that in some time-slot, a node may not have any
neighbors.

\section{Scheme with Constant Throughput Scaling}
\label{s:scheme_Tput}

In this section we present Scheme $\Pi$ and show that it achieves
constant throughput scaling. In the next section its delay scaling
will be analyzed. Before presenting the scheme, we prove a property of
the random network model which makes the scheme feasible.

\begin{lemma}
Configurations are typical {\em whp}.
\label{l:typical_whp} 
\end{lemma}
\begin{proof}
Consider any two nodes $i$ and $j$. First note that the probability
that $G_i$ and $G_j$ coincide is zero. Also any two distinct great
circles necessarily intersect at exactly two points. By definition,
${\mathcal C}_{ij}$ has area $c_1/n$ since it has radius
$(2+\Delta)\sqrt{\pi/n}$.

Let $I_k, ~k=1,\ldots,n, ~k \neq i,j,$ be an indicator random variable
for the event that the great circle of node $k$, $G_k$, passes through
${\cal C}_{ij}$. By definition, $I_k$ are i.i.d. Bernoulli random
variables with parameter $p$, where $p = c_2/\sqrt{n}$ where $c_2$ is
a positive constant.  This is because a great circle passes through a
disk of radius $R$ if and only if its pole lies in an equatorial band
of width $2R$. The probability of this event is $\Theta(R)$ as the
position of pole is uniformly distributed over the sphere.

Thus, the total number of great circles passing through ${\cal
C}_{ij}$ is given by a random variable $X = \sum_{k} I_k$ with $E[X]
\approx 0.5c_1\sqrt{n}=\Theta\left(\sqrt{n}\right)$. An application of
the well-known Chernoff bound for the sum of i.i.d. Bernoulli random
variable (e.g., see \cite{MR95}), yields
\begin{eqnarray}
P\{| X - E[X] | \geq \delta E[X]\}
& \leq &2\exp\left(-\delta^2E[X]/2\right) \nonumber \\ 
& = & \frac{1}{n^3},~~~
\mbox{for $\delta = \sqrt{\frac{2(\log 2+3\log n)}{E[X]}}$}. 
\label{e:may1}
\end{eqnarray}

The choice of $\delta$ in (\ref{e:may1}) shows that $X \leq c_1 \sqrt{n}$  
or $X = \Theta\left(\sqrt{n}\right)$ with probability at least
$1-1/n^3$. Hence by the union bound over all $n(n-1)/2$ possible
${\cal C}_{ij}$ for $i,j=1,\ldots,n$, we obtain that with probability
at least $1-1/n$, the number of great circles passing through each
${\cal C}_{ij}$ is $\Theta\left(\sqrt{n}\right)$.
\end{proof}

\subsection{The Scheme}

The operation of Scheme $\Pi$ depends on whether the configuration is
typical or not. If the configuration is not typical, direct
transmission is used between the S-D pairs along with time-division
multiplexing. That is, the sources transmit to their destinations once
in $2/n$ time-slots in a round-robin fashion. If the configuration is
typical then Policy $\Sigma_n$ as described below is used. Policy
$\Sigma_n$ is a variant of the policies presented in \cite{GT},
\cite{DGT}.

\begin{figure}[htpb]
\begin{center}
\begin{psfrags}
\psfrag{S}[l]{\rm S}
\psfrag{R}[b]{\rm R}
\psfrag{D}[r]{\rm D}
\psfrag{Csr}[r]{${\cal C}_{SR}$}
\psfrag{Crd}[l]{${\cal C}_{RD}$}
\includegraphics[width=3in,angle=0]{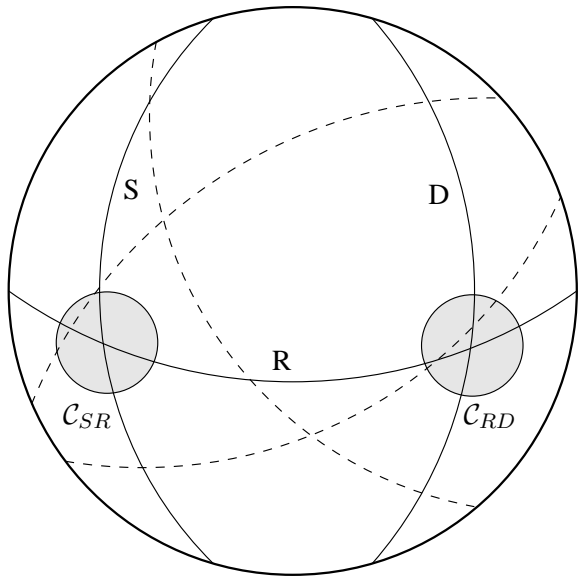}
\end{psfrags} 
\caption{Source node, S, transmits its packet to a random relay node, R. 
The packet is carried by R, until its transmission to the destination
node, D, when R and D become neighbors. The dotted great circles
correspond to other nodes which can act as relays.}
\label{f:srdGreatCircles}
\end{center}
\end{figure}

\medskip \hrule \medskip
\noindent{\bf Policy $\Sigma_n $:}
\medskip \hrule \medskip
\begin{enumerate}
\item Each time-slot is divided into two sub-slots -- A and B.

\item Sub-slot A
\begin{enumerate}
\item[(a)] Each source node independently becomes {\em active} with probability 
$p_{\Delta}>0$.

\item[(b)] If an active node has one or more neighbors then with probability
$0<\alpha<1$, it chooses one at random and a packet intended for its
destination is transmitted to this randomly chosen neighbor, which
acts as a relay node.
\end{enumerate}

\item Sub-slot B
\begin{enumerate}
\item Each node independently becomes {\em active} with probability 
$p_{\Delta}>0$.

\item If an active node has one or more neighbors that are destination nodes, 
it chooses one at random.  The active node, which acts as a relay,
transmits a packet intended for this destination node, if it has any,
in FIFO order.
\end{enumerate}

\end{enumerate}
\medskip \hrule \medskip
In policy $\Sigma_n$, each node acts as a relay for all the other $n/2-1$
S-D pairs. A packet reaches from its source to its destination as
shown in Figure~\ref{f:srdGreatCircles}. A source node, S, transmits
its packet to a random relay node, R, which may also happen to be the
destination itself. The random relay node then moves around carrying
the packet. Finally, when it becomes a neighbor of the destination, D,
the packet is transmitted to D. A relay node may receive several
packets from a source before it gets a chance to transmit to the
destination. To handle this, each relay node maintains a separate
queue for each of the other $n/2 -1$ S-D pairs.

The actual mechanism is slightly more complicated. Since each node
decides to transmit at random, it is possible that two nearby nodes
transmit simultaneously so that transmission is not successful under
the Protocol model. In order to analyze the throughput of Scheme
$\Pi$, we first state a result about the probability of successful
transmission between two nodes when they are neighbors under policy
$\Sigma_n$.

\begin{lemma}
\label{l:tput1} Under policy $\Sigma_n$, the following hold in a typical 
configuration.
\begin{enumerate}
\item[(a)] In sub-slot A, if nodes S and R are neighbors of each other, S
transmits a packet to R successfully with a strictly positive
probability, independent of $n$.
\item[(b)] In sub-slot B, if nodes R and D are neighbors of each other, R
transmits a packet to D successfully with a strictly positive
probability, independent of $n$.
\end{enumerate}
\end{lemma}
\begin{proof}
We shall only prove for the case of sub-slot A since the proof for the
other part is similar. Consider a sub-slot A in which S and R are
neighbors. Let $E_1$ be the event that S becomes active and $E_2$ be
the event that S chooses R as a random relay and no other source node
in ${\cal C}_{SR}$ becomes active. If both events $E_1$ and
$E_2$ occur, S transmits to R and the transmission is successful under
the Relaxed Protocol model. Thus,
\begin{eqnarray}
P( \mbox{ S transmits to R successfully}) & = & P(E_1 \cap E_2)
\nonumber \\ & = & P(E_1) P(E_2|E_1).\label{e:SRsuccess_prob}
\end{eqnarray}

From the description of Policy $\Sigma_n$ it is clear that
$P(E_1)=\alpha p_\Delta$, which is a strictly positive constant. Next we
compute $P(E_2|E_1)$ and show that it is lower bounded by a strictly
positive constant, independent of $n$, which will imply the statement
of the lemma.

Given that S is active, the probability of successful transmission to
R depends on how many other nodes are present in ${\cal C}_{SR}$ since
these nodes could interfere, i.e., transmit simultaneously so that the
transmission from S to R is not successful under the Relaxed Protocol
model.

Since we have a typical configuration, $\Theta(\sqrt{n})$ distinct
great circles or source nodes intersect ${\cal C}_{SR}$. Moreover each
great circle has $\Theta(1)$ lattice points that are in ${\cal
C}_{SR}$. For a natural random walk on a discrete torus of size
$\sqrt{n}$, the probability of being at any particular position is
$1/\sqrt{n}$. Hence the probability that any of the
$\Theta\left(\sqrt{n}\right)$ source nodes whose great circles
intersect ${\mathcal C}_{SR}$ is present in ${\cal
C}_{SR}$ with probability $\Theta(1/\sqrt{n})$. Due to the independent
movement of all nodes, we obtain that for a typical configuration, the
probability of $k$ nodes being present in the ${\cal C}_{SR}$ is
\[q(k) = {c_1\sqrt{n} \choose k} \left(\frac{c_3}{\sqrt{n}}\right)^k
\left(1-\frac{c_4}{\sqrt{n}}\right)^{c_2\sqrt{n}-k} \approx
\frac{\left(c_1c_3\right)^k \exp(-c_2c_4)}{k!},\] 
for large enough $n$. If ${\cal C}_{SR}$ has $k$ nodes not including S
and R then S certainly has no more than $k+1$ neighbors. In this
situation, R is chosen by S with probability at least
$1/(k+1)$. Further there are at most $k$ other source nodes and the
probability that no other node in ${\cal C}_{SR}$ becomes active is at
least $(1-p_{\Delta})^k$. Thus,
\begin{eqnarray*}
P(E_2|E_1) & \geq & \sum_{k = 0}^{n-2} \frac{\left(c_1c_3\right)^k
\exp(-c_2c_4)}{k!} \frac{1}{k+1} (1-p_\Delta)^k\nonumber\\ 
&\geq& \exp(-c_2c_4)\sum_{k = 0}^{n-2} \frac{\left(c_1c_3(1-p_\Delta)
\right)^k}{(k+1)!}.
\end{eqnarray*}
It is easy to see that for $0 < p_{\Delta} < 1$, the term on the right
hand side is a lower bounded by a strictly positive constant. Hence,
$P(E_2|E_1)$ is strictly positive. This completes the proof of the
lemma.
\end{proof}

\begin{theorem}
Scheme $\Pi$ achieves  $T(n)=\Theta(1)$.
\label{t:schemepi_Tput} 
\end{theorem}
\begin{proof}
Consider a typical configuration so that policy $\Sigma_n$ is
used. Fix a source node S and a relay node R. Let $A(t)$ be the number
of bits transmitted from S to R in sub-slot A of time-slot $t$. If S
transmits to R successfully in sub-slot A of time-slot $t$, $A(t)=W/2$
otherwise $A(t)=0$.

First we determine $E[A(t)]$. Let $F_1$ be the event that S and R are
neighbors and $F_2$ be the event that S transmits to R
successfully. Then
\begin{equation}
E[A(t)] = \frac{W}{2}P\{F_1\cap F_2\} = \frac{W}{2}P\{F_1\}P\{F_2|F_1\}.
\label{e:SRtput}
\end{equation}
From Lemma~\ref{l:tput1}(a), $P\{F_2|F_1\} \geq c_5 > 0$. Due to the
independent motion of nodes S and R according to natural random walks,
the joint description of their positions is a two-dimensional random
walk on a discrete torus of size $\sqrt{n}\times\sqrt{n}$. It is easy
to see that the stationary distribution for this process is the
uniform distribution on $n$ joint positions. Since S and R become
neighbors when they are in one particular joint position out of these
$n$ joint positions, it follows that the probability of S and R being
neighbors is $1/n$, i.e., $P(F_1)=1/n$. Hence from (\ref{e:SRtput}) it
follows that $E[A(t)]=\Theta(1/n)$.

Now the positions of nodes S and R form an irreducible, finite state
Markov chain and $A(t)$ is a bounded, non-negative function of the
state of this Markov chain at time $t$. Therefore by the ergodicity of
such a Markov chain, the long-term throughput between S and R is
\[\lim_{T\rightarrow\infty} \frac{1}{T}\sum_{t=1}^T A(t) = E[A(t)] =\Theta(1/n).\]
Thus the throughput between a source node S and any other node in
sub-slot A is $\Theta(1/n)$. Similarly, it can be shown that the
throughput between any node and a destination node D in sub-slot B is
also $\Theta(1/n)$. The value of $0<\alpha<1$ guarantees that the
arrival rate of packets belonging to every S-D pair at any relay node
is strictly less than the service rate. This ensures the stability of
the queues formed at the relay nodes, which in turn implies that the
throughput between each S-D pair is simply the sum of the throughputs
between S and the other $n-1$ nodes in sub-slot A. Hence the
throughput of each S-D pair is $\Theta(1)$.

We have shown that in a typical configuration, Scheme $\Pi$ provides
$\Theta(1)$ throughput between all S-D pairs. From
Lemma~\ref{l:typical_whp}, configurations are typical {\em whp}. Hence
it follows that Scheme $\Pi$ has throughput $T(n)=\Theta(1)$.
\end{proof}

Note that for the unrestricted mobility models in \cite{GT} and
\cite{EMPS_TDto_fluid}, it is possible to prove a stronger result that each
S-D pair has $\Theta(1)$ throuhgput for any $n$ with probability $1$
instead of probability approaching $1$ as $n$ tends to infinity, as in
the present case.

\section{Delay of Scheme $\Pi$} 
\label{s:delay}
Under Scheme $\Pi$, if the configuration is not typical, direct
transmission is used, in which case the delay for each packet is
$1$. Since the delay of a scheme is defined to be the expectation over
all configurations of the average delay, the delay for Scheme $\Pi$ is
determined by the expected delay over typical configurations. So we
shall assume that the configuration is typical.

Consider a particular S-D pair. Packets from S reach D either directly
by a single hop in sub-slot A or through any of the other $n-2$ nodes,
which act as relays. Since the nodes perform independent random walks,
only $\Theta(1/n)$ of the packets belonging to any S-D pair reach
their destination in a single hop. Thus, most of the packets reach
their destination via a relay node, in which case the delay is two
time-slots for two hops plus the mobile-delay, which is the time spent
by the packet at the relay node.

Each relay node maintains a separate queue for each of the S-D
pairs. Fix a relay node, R, and consider the queue for the S-D pair
under consideration. The mobile-delay mentioned above is the delay at
this relay-queue.To compute the average delay for this relay-queue, we
need to study the characteristics of its arrival and potential
departure processes.

First we obtain a lower bound on the delay at the relay-queue. Each
node performs a random walk on a 1-D torus of size $\sqrt{n}$ on its
great circle. We say that an S-D pair intersects node R's great circle
$k$ vertices apart if the lattice points where R can become neighbors
of S and D are $k$ lattice points (vertices) apart on the 1-D discrete
torus of R.

Fix an S-D pair and consider a particular relay node R. When a packet
is transmitted successfully from S to R, D is equally likely to be in
any of its $\sqrt{n}$ lattice points since it performs an independent
random walk. Let $T_{ij}$ be the random time it takes for a random
walk on a $\sqrt{n}\times\sqrt{n}$ torus to hit $(0,0)$ starting from
$(i,j)$. If the S-D pair intersects the great circle of R $i$ vertices
apart then the expected delay for packets of this S-D pair relayed
through R is lower bounded by $\sum_{j=0}^{\sqrt{n}-1} T_{ij}$.

Using the Chernoff bound for the sum of i.i.d. Bernoulli random
variable (e.g., see \cite{MR95}), it can be shown that
$\Theta(\sqrt{n})$ S-D pairs intersect the great circle of each node
$i$ points apart for $0\leq i \leq \sqrt{n}-1$ {\em whp}. Hence the
delay of Scheme $\Pi$, which is the expected delay over all packets is
\[D(n) = \Omega\left(E\left[\frac{1}{n}\sum_{i,j=1}^{\sqrt{n}-1} T_{ij} \right]\right).\]
As shown in \cite{AF94}, $E\left[\frac{1}{n}\sum_{i,j=1}^{\sqrt{n}-1}
T_{ij} \right] = \Theta(n\log n)$. Therefore,
\begin{equation}
D(n)=\Omega(n\log n).
\label{e:LB_delay}
\end{equation}

The rest of this section derives an upper bound which is of the same
order as the lower bound. It is hard to obtain an upper bound on the
delay in the relay-queue since the arrival and service processes are
complicated and dependent. We progressively obtain queues that are
simpler to analyze and upper bound the delay of the previous queue as
follows. We first upper bound the delay in the relay-queue by that in
another queue, ${\cal Q}_1$, in which the arrival process is
simpler. The delay of ${\cal Q}_1$ is upper bounded by that in ${\cal
Q}_2$, which has a relatively simpler service process. However, the
arrival and service process are not independent. The final part
consists of introducing a virtual server with i.i.d. Geometric service
times to break this dependence. With this overview, we proceed to the
details.

Recall that a packet arrives at the relay-queue when (i) S and R are
neighbors, (ii) S becomes active (which happens with probability
$\alpha p_\Delta$), (iii) S chooses R as a random relay, and (iv) the
transmission from S to R is successful. Similarly, a packet can depart
from the queue when (i') R and D are neighbors, (ii') R becomes active
(which happens with probability $p_\Delta$), (iii') R chooses D as the
destination node, and (iv') the transmission is successful. We call
such a time-slot a potential departure instant and the sequence of
inter-potential-departure times is called the potential-departure
process. Let the potential-departure process of the relay-queue be
called $\{S_i\}$. The qualifier potential is used since a departure
can occur only if R has a packet for D.

Consider a queue ${\cal Q}_1$ in which arrivals happen whenever (i),
(ii) and (iii) above are satisfied, irrespective of whether (iv) is
satisfied or not. The potential departure process for ${\cal Q}_1$ is
the same as that for the relay-queue. Then it is clear that the
expected delay in ${\cal Q}_1$ provides an upper bound on that in the
relay-queue.

Recall that the motion of each node is an independent 1-D random walk
on a discrete torus of size $\sqrt{n}$. We will say that two nodes
{\em meet} when they become neighbors. Since nodes move independently
the joint position of nodes R and D is a random walk on a
$\sqrt{n}\times\sqrt{n}$ discrete torus and R and D become neighbors
when the 2-D random walk is in state $(0,0)$, without loss of
generality. Therefore, the inter-meeting time of R and D is
distributed like the inter-visit time of state $(0,0)$ of a 2-D random
walk. Since this is a Markov chain with $n$ states having a uniform
stationary distribution, we know that the sequence of inter-meeting
times of nodes R and D, denoted by $\{\tau_i, i\geq 0\}$, is an
i.i.d. process. Further, if $\tau$ is a random variable with the
common distribution then
\begin{equation}
E[\tau] = n.
\label{e:Etau}
\end{equation}

However a potential departure instant does not occur each time R and D
meet. A potential departure instant occurs only if R also becomes
active, chooses D as the random destination and the transmission is
successful. If R and D are not chosen in spite of being in the same
cell, it increases the likelihood of there being many more nodes in
the same cell. Due to the random walk model of the node mobility, if
there is a crowding of nodes in some part of the network then it
remains crowded for some time in the future. Hence due to the
Markovian nature of node mobility, the inter-potential-departure times
are not independent.

We want to obtain an upper bound on the delay of ${\cal Q}_1$ which
has potential-departure process $\{S_i\}$. To do this we will consider
a queue, ${\cal Q}_2$, which has the same arrival process as ${\cal
Q}_1$ but a different departure process $\{\tilde{S}_i\}$ such that
$S_i \leq \tilde{S}_i$. Then the expected delay in ${\cal Q}_2$ would
provide an upper bound on the the expected delay in the relay-queue.

\begin{figure}[htpb]
\begin{center}
\begin{psfrags}
\psfrag{R}[l]{\rm R}
\psfrag{D}[b]{\rm D}
\psfrag{A}[b]{$A$}
\psfrag{0}[tr]{$0$}
\psfrag{fn}[r]{$m$}
\psfrag{fn1}[t]{$\frac{\zeta}{2}$}
\psfrag{fn2}[t]{$\frac{\zeta}{2}+1$}
\psfrag{fn3}[t]{$m$}
\includegraphics[width=3in,angle=0]{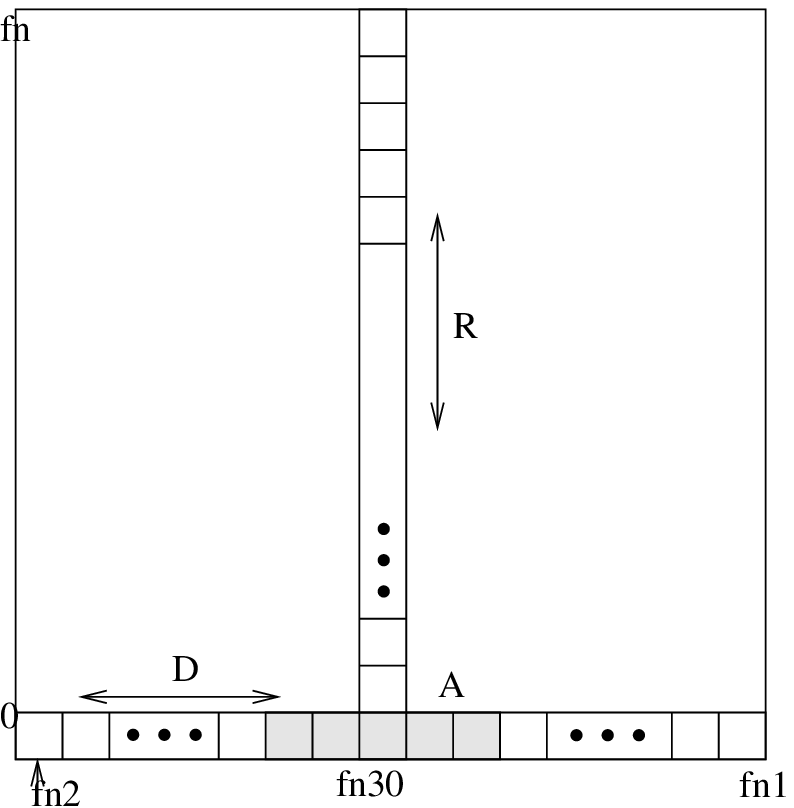}
\end{psfrags} 
\caption{Schematic representation of the motion of nodes R and D on their 
respective great circles with $\zeta=\sqrt{n}-1$.}
\label{f:RDrw}
\end{center}
\end{figure}

Nodes R and D perform independent random walks on 1-D tori of size
$\sqrt{n}$ on their great circles as shown in
Figure~\ref{f:srdGreatCircles} and R and D meet when both are at a
particular pair of lattice points. This is represented schematically
in Figure~\ref{f:RDrw}, where R performs a vertical 1-D random walk
and D performs a horizontal 1-D random walk. The joint motion of nodes
R and D is equivalent to a random walk on a 2-D torus of size
$\sqrt{n}\times\sqrt{n}$ and R and D meet when this 2-D random walk is
in state $(0,0)$. The inter-meeting times of nodes R and D correspond
to the i.i.d. process $\{ \tau_i \}$.  Further, let $\alpha_i = \tau_1
+ \ldots + \tau_i$ for $i\geq 1$, i.e., $\alpha_i$ is the time-slot in
which R and D meet for the $i$th time.  In a typical configuration, we
know that the number of other great circles that pass through ${\cal
C}_{RD}$ is $\Theta\left(\sqrt{n}\right)$. Allowing for the worst
case, based on Lemma \ref{l:typical_whp}, let there be
$c_1\sqrt{n}=m-2$ other great circles that pass through ${\cal
C}_{RD}$. These can also be thought of as performing independent
random walks on the horizontal 1-D torus. Let nodes R and D be
numbered $1$ and $2$ and the other $c_1\sqrt{n}$ nodes be numbered
from $3$ to $m$ and let $X(t) = (X_1(t),\dots,X_m(t))$ denote the
position of these $m$ nodes on the $\sqrt{n}\times\sqrt{n}$ discrete
torus at time $t$.

\begin{figure}[htpb]
\begin{center}
\begin{psfrags}
\psfrag{t1}[b]{$\tau_1$}
\psfrag{t2}[b]{$\tau_2$}
\psfrag{t3}[b]{$\tau_3$}
\psfrag{t4}[b]{$\tau_4$}
\psfrag{t5}[b]{$\tau_5$}
\psfrag{s1}[t]{$S_1$}
\psfrag{s2}[t]{$S_2$}
\psfrag{s3}[t]{$S_3$}
\psfrag{e10}[c]{$E_1=0$}
\psfrag{e21}[c]{$E_2=1$}
\psfrag{e31}[c]{$E_3=1$}
\psfrag{e40}[c]{$E_4=0$}
\psfrag{e51}[c]{$E_5=1$}
\includegraphics[width=5in,angle=0]{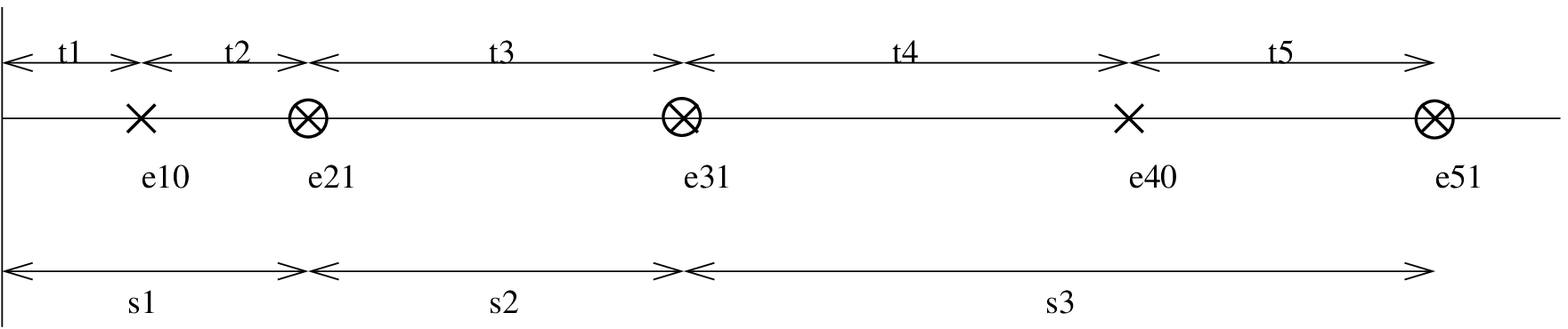}
\end{psfrags} 
\caption{ The `x' marks correspond to the times when R and D meet 
each other. At some of these meeting instants the R-D transmission can
be successful. Such points have been circled and correspond to
$E_i=1$. The inter-potential-service times are thus the sum of a few
of the inter-meeting times of R and D.}
\label{f:generateSi}
\end{center}
\end{figure}

A constant number of lattice points of the 1-D torus correspond to
${\cal C}_{RD}$ and these are shown by the shaded region in
Figure~\ref{f:RDrw} and is referred to as set $A$. Let $E_i$ be the
indicator for the event that R chooses D and the transmission is
successful in time-slot $\alpha_i$. That is, $E_i$ is the indicator
for the event that $\alpha_i$ is a potential departure instant. Let
$N_i$ be the number of other destination nodes in $A$ in time-slot
$\alpha_i$. Then $P\{E_i=1\}$ depends on $N_i$ only. Now, $N_i$ 
depends on   $X(\alpha_i)$ which depends
on the past given by $E^{i-1} = \{E_0,\ldots,E_{i-1} \}$ and $\tau^i =
\{\tau_0, \ldots, \tau_i \}$. Thus the potential-departure process is
generated by choosing some of the meeting instants of R and D
according to a probability modulated by $N_i$, which is another
independent process as shown in Figure~\ref{f:generateSi}.

Above we described how the process $\{S_i\}$ can be generated using
the processes $\{N_i\}$ and $\{\tau_i\}$, which in turn were obtained
from $\{X(t)\}$, which corresponds to the independent random walks of
all $m$ nodes. Next we shall perturb the process $\{X(t)\}$ to obtain
$\{\tilde{X}(t)\}$ and the corresponding $\{\tilde{\tau}_i\}$ and
$\{\tilde{N}_i\}$. Let $Z(t)$ be a 1-D horizontal random walk on a
torus of size $\sqrt{n}$. Let $\tilde{X}_i(t) = X_i(t) + Z(t)$ be the
position of node $i, 1\leq i\leq m$, where the addition is modulo
$\sqrt{n}$. Then the inter-meeting times of any two nodes are the same
as before since the position of each node is shifted horizontally by
the same amount due to $Z(t)$. As a result the processes $\tau_i$ and
$\tilde{\tau}_i$ are identical.  Under the modified setup, the lattice
point at which R and D meet can be any element of the set
$B=\{(i,0):0\leq i\leq \sqrt{n}-1\}$ instead of always being $(0,0)$.
Similarly, let $\tilde{N}_i$ be the number of other destination nodes
in the set $A+Z(t)$. Then, $\{\tilde{N}_i\}$ is identical to
$\{N_i\}$. Thus the process $\{S_i\}$ can also be generated (through
$\{E_i\}$) using $\{\tilde{N}_i\}$ and $\{\tilde{\tau}_i\}$ instead of
$\{N_i\}$ and $\{\tau_i\}$.  Therefore we shall use $\tilde{X}_i(t)$
as the position of node $i$ at time $t$ instead of $X_i(t)$. Under
this perturbed motion, R can be seen as if it performs a 2-D random
walk on the $\sqrt{n}\times\sqrt{n}$ torus while D and the other $m-2$
nodes perform a 1-D random walk on a 1-D torus of size $\sqrt{n}$
which is subset $B$ of the 2-D torus. Moreover, given
$\tilde{X}_3^m(\alpha_i) =
(\tilde{X}_3(\alpha_i),\ldots,\tilde{X}_m(\alpha_i))$, $P\{E_i=1\}$ is
independent of everything else.

\begin{lemma}
There exists a constant (independent of $n$) $c_6 > 0$ such that
\[P\left(E_i=1|\tau^i, E^{i-1}\right) \geq c_6 > 0.\]
\label{l:LB_Pmark}
\end{lemma}
\begin{proof}
The initial position of R, $X_1(0)$ has a uniform distribution of the
$\sqrt{n}\times\sqrt{n}$ torus. The initial positions of D and nodes
$3$ to $m$ have independent uniform distributions on subset $B =
\{(i,0):0\leq i\leq \sqrt{n}-1 \}$ of the $\sqrt{n} \times \sqrt{n}$ 
torus. As a result $X_1(\alpha_1) = X_2(\alpha_1) = I$ where $I$ is a
random variable with a uniform distribution over $B$.

Let $V=(\tilde{X}_3(\alpha_i), \ldots, \tilde{X}_m(\alpha_i))$ be the
configuration of the $m-2$ nodes other than R and D. Then the
conditional probability of a potential departure given the past can be
written as
\begin{eqnarray}
P\left(E_i=1|\tau^i,E^{i-1}\right)
&=&\sum_V \frac{P\left(E_i=1|V,\tau^i,E^{i-1}\right)P\left(V, \tau^i, E^{i-1}\right)}
{P(\tau^i,E^{i-1})}\nonumber\\ 
& \geq & \min_{V} P(E_i=1|V,\tau^i,E^{i-1}) \left[ \sum_{V} 
\frac{P(V,\tau^i,E^{i-1})}{P(\tau^i,E^{i-1})} \right] \nonumber\\
&= & \min_{V} P(E_i=1|V,\tau^i,E^{i-1}) \nonumber \\ & = & \min_{V} P\left( E_i=1 | V \right), 
\label{e:LB_condPE1}
\end{eqnarray}
where the last equality holds because $E_i$ is independent of
everything else given $V$.

Given a congfiguration $V$, the number of nodes in $A+(i-1,0)$ for
$i=1,\ldots,\sqrt{n}$ torus can be found and this in turn determines
the $P\left(E_i=1|V\right)$. Hence, if $V_i$ denotes the number of
nodes other than R and D in the set $A+(i-1,0)$ for
$i=1,\ldots,\sqrt{n}$ then we can equivalently let the configuration
be $V=(V_1,\dots,V_{\sqrt{n}})$. 

Now consider a fixed configuration, $V=v=(v_1,\ldots,v_{\sqrt{n}})$,
and let $Z$ be a random variable which takes value $v_i, 1\leq i\leq
\sqrt{n}$ with probability $1/\sqrt{n}$. Let $A$ consists of $c_2$
(some constant) elements. Then
\begin{equation}
E[Z] = \frac{1}{\sqrt{n}} \sum_{k=1}^{\sqrt{n}} v_k = \frac{c_2(m-2)}{\sqrt{n}} = \Theta(1).
\label{e:EZ}
\end{equation}

Recall that $X_1(\alpha_i)=I$, where $I$ is a random variable with
uniform distribution on $B$. Further, from the description of Scheme
$\Pi$, if there are $v_k$ destination nodes other than D in ${\cal
C}_{RD}$ then $E_i=1$ if R chooses D out of all destination nodes that
are its neighbors and the other $v_k$ nodes do not transmit. Since
${\cal C}_{RD}$ contains all neighbors and more, the number of
neighbors can be no more than $X_i$ and hence for $k=1, \ldots,
\sqrt{n}$, we obtain
\begin{equation}
P(E_i=1 | V=v, X_1(\alpha_i)=(k-1,0)) \geq \frac{p_\Delta (1-p_\Delta)^{v_k + 1}}{v_k + 1}.
\label{e:PE1cond_occupancy}
\end{equation}

Define a real valued function $f: {\mathbb R} \to {\mathbb R}$ where
$f(x) = \frac{p_{\Delta}(1-p_{\Delta})^{x+1}}{x+1}$.  It is easy to
check that $f(\cdot)$ is a convex function. Hence, by Jensen's
inequality,
\begin{eqnarray}
E[f(Z)] & \geq & f(E[Z]). 
\label{e:fjensen}
\end{eqnarray}
Using (\ref{e:EZ}), (\ref{e:PE1cond_occupancy}) and (\ref{e:fjensen}),
for any configuration $V$ with corresponding $v$, we obtain
\begin{eqnarray}
P(E_i=1|V=v) &=& \sum_{k=1}^{\sqrt{n}} P( E_i=1|V=v,X_1(\alpha_i)=(k-1,0)) P(X_1(\alpha_i)=(k-1,0)|V=v) \nonumber\\
&=& \frac{1}{\sqrt{n}} \sum_{k=1}^{\sqrt{n}} P (E_i=1|V=v,X_1(\alpha_i)=(k-1,0)) \nonumber \\
&\geq& \frac{1}{\sqrt{n}} \sum_{k=1}^{\sqrt{n}}  \frac{p_\Delta (1-p_\Delta)^{v_k + 1}}{v_k + 1} \nonumber \\
&=& E[f(Z)] ~ \geq ~ f(E[Z]) \nonumber \\
&=& f\left(\frac{c_2(m-2)}{\sqrt{n}}\right) \stackrel{\triangle}{=}
c_6>0. \label{e:f} 
\end{eqnarray} 
Combining (\ref{e:LB_condPE1}) and (\ref{e:f}) completes the proof of
the lemma.
\end{proof}

Recall that the process $\{S_i\}$ is generated from $\{\tau_i\}$ and
$\{E_i\}$. Consider an i.i.d. Bernoulli process $\{\tilde{E}_i\}$ with
$P\{\tilde{E}_1=1\}=c_6$. Now we can construct a process
$\{\tilde{S}_i, i\geq 1\}$ similar to the process $\{S_i\}$ using
$\{\tau_i\}$ and $\{\tilde{E}_i\}$ instead of $\{E_i\}$.
Lemma~\ref{l:LB_Pmark} shows that the processes $\{S_i\}$ and
$\{\tilde{S}_i\}$ are coupled such that $\tilde{S}_i \geq S_i$ (the
inequality corresponds to standard stochastic dominance).  Now
consider queue, ${\cal Q}_2$, with the same arrival process as ${\cal
Q}_1$ but with potential-departure process
$\{\tilde{S}_i\}$. Depending on the value of $c_6$, the value of
$\alpha$ can be chosen so that the arrival rate is strictly smaller
than the potential departure rate in ${\cal Q}_2$ so as to ensure
stability. The distribution of $\tilde{S}_1$ is the same as $\tau_1 +
\ldots + \tau_G$, where $G$ is an independent Geometric random
variable with parameter $c_6$. As a result, for any $r \in {\mathbb
N}$,
\begin{equation}
E[\tilde{S}^r_1]=\Theta(E[\tau^r_1]).
\label{e:EStilde2}
\end{equation}
In light of (\ref{e:EStilde2}), it is easy to see that the delay
scaling of queue $\cq_2$ is the same as the delay scaling of a queue
in which an arrival happens each time S and R meet with probability
$0.5$ and a potential departure occurs each time R and D meet. Since
we are interested only in the delay scaling, henceforth we assume that
in $\cq_2$, an arrival happens when S and R meet with probability
$0.5$ and a potential departure occurs whenever R and D meet.

At this stage we have upper bounded the delay in the relay-queue by
the delay in ${\cal Q}_2$. The inter-arrival times and the
inter-potential departure times in ${\cal Q}_2$ are i.i.d. processes.
However these two processes are not independent for the following
simple reason:~if the S-D pair intersects the great circle of R, $k >
0$ vertices apart then R has to travel at least distance $k$ on the
discrete torus after an arrival for a potential departure to occur.

Next, we will bound the delay in $\cq_2$ by the sum of the delays
through two {\em virtual} queues, $\cq_3$ and $\cq_4$, in tandem. Both
$\cq_3$ and $\cq_4$ will be shown to have delay of $O(n \log n)$. This
will imply that the delay of $\cq_2$ is $O(n\log n)$. Queues $\cq_3$
and $\cq_4$ are constructed as follows. The arrival process of $\cq_3$
is the same as that of $\cq_2$. The potential-departure process of
$\cq_3$ is an i.i.d. Bernoulli process with parameter $2/3n$ (or
potential departure rate $\frac{2}{3n}$). An arrival occurs at $\cq_4$
whenever there is a potential-departure at $\cq_3$. If $\cq_3$ is
non-empty, then the arrival to $\cq_4$ is the head-of-line packet
transferred from $\cq_3$ to $\cq_4$ or else a {\em dummy} packet is
fed to $\cq_4$. Thus the arrival process at $\cq_4$ is the same as the
potential-service process at $\cq_3$. By construction, the delay of a
packet through this tandem of queues, $\cq_3$ and $\cq_4$, upper
bounds the delay experienced by a packet through $\cq_2$. Now, from
Lemmas~\ref{lx1} and \ref{lx2} stated in the next section, the
expected delay through $\cq_3$ and $\cq_4$ is $O(n\log n)$. Thus the
expected delay of the packets of each S-D pair relayed through each
relay R in a typical configuration is $O(n\log n)$. The delay of
scheme is the expectation of the packet delay averaged over all S-D
pairs and all relay nodes. Hence it follows that the delay of the
scheme is $O(n\log n)$. Combining this with (\ref{e:LB_delay}), we have
proved the following.

\begin{theorem}
The delay of Scheme $\Pi$ is $\Theta(n\log n)$.
\label{t:schemepi_delay}
\end{theorem}

\section{Remaining Proofs}
\label{s:rem_proofs}
In this section, we prove Lemmas \ref{lx1} and \ref{lx2}, which were
used to prove that Scheme $\Pi$ has delay of $O(n\log n)$. Before
proving these, we present Lemma \ref{lem:var} which will be useful for
both these proofs.

Recall that each node performs an independent random walk on a 1-D
discrete torus of size $\sqrt{n}$ on its great circle. Let $Z$ be a
random variable which is distributed as the inter-meeting time of two
distinct nodes. The following lemma provides the first two moments of
$Z$.
\begin{lemma}
\label{lem:var}
\[ E[Z] = n, ~~~E[Z^2] = \Theta(n^2 \log n).\]
\end{lemma}
\begin{proof}
As nodes perform independent random walks, the joint position of two
nodes is a 2-D random walk on the $\sqrt{n}\times\sqrt{n}$ discrete
torus. Thus the inter-meeting time of any two nodes is equivalent to
the first return time to state $(0,0)$ for this random walk on a
$\sqrt{n}\times\sqrt{n}$ torus. Since we are interested only in
determining the exact order of the moments, we will consider a simple
random walk.

Let $X(t) = (X_1(t),X_2(t)) \in \{0,\dots,\sqrt{n}-1\}^2$ be a simple
random walk on the $\sqrt{n}\times\sqrt{n}$ torus. Then the first
return time to state $(0,0)$ is
\[ T=\inf\{ t \geq 1:X(t) = (0,0),\; X(0)=(0,0)\}. \]
Note that $X(t)$ is a finite-state Markov chain with a uniform
equilibrium on the $n$ states. For any finite-state Markov chain, the
expectation of the first return time to any state is the reciprocal of
the equilibrium probability of the Markov chain being in that
state. Hence, $E[T] = n$. 

Define, $T_0 = \inf\{ t \geq 1: X(t) = (0,0)\}$. Observe that $T_0$
differs from $T$ in that $T$ is conditioned on starting at
$X(0)=(0,0)$.  Let $E_{(i,j)}T_{(k,l)}$ denote the expected time to
hit state $(k,l)$ for the first time starting from state $(i,j)$. Let
$E_\pi[T_0]$ denote the expectation of $T_0$ given that $X(0)$ is
distributed according to the uniform stationary probability
distribution $\pi$. Then,
\begin{eqnarray}
E_\pi[T_0]&=&\sum_{i,j=0}^{\sqrt{n}-1}\pi(i,j)E_{(i,j)}T_{(0,0)}
~=~ \sum_{i,j=0}^{\sqrt{n}-1}\pi(i,j)E_{(i,j)}T_{(k,l)}
\label{e:t0_1}\\ 
&=& \sum_{i,j=0}^{\sqrt{n}-1}\sum_{k,l=0}^{\sqrt{n}-1}\frac{1}{n}\pi(i,j)E_{(i,j)}T_{(k,l)}~=~\sum_{i,j=0}^{\sqrt{n}-1}\sum_{k,l=0}^{\sqrt{n}-1}\pi(i,j)\pi(k,l)E_{(i,j)}T_{(k,l)}\nonumber\\
&=&n\log n,
\label{e:t0_2}
\end{eqnarray}
where (\ref{e:t0_1}) holds because $\sum_{ij} E_{(i,j)}T_{(0,0)} =
\sum_{ij} E_{(i,j)}T_{(k,l)}$ for any $0\leq k,l\leq \sqrt{n}-1$ due to symmetry
of states corresponding to cells on the torus. For the validity of
(\ref{e:t0_2}), see page 11 of Chapter 5 in \cite{AF94}.

Using Kac's formula (see Corollary 24 in Chapter 2 of \cite{AF94}) and
(\ref{e:t0_2}), we obtain
\begin{eqnarray}
E[T^2] &=& \frac{2E_\pi[T_0] + 1}{\pi(0,0)}~=~ 2n^2\log n + n.\nonumber\\
\end{eqnarray}
Therefore, we obtain $E[Z] = n$ and $E[Z^2] = \Theta(n^2 \log n)$.
\end{proof}

\begin{lemma}\label{lx1}
Let $D_3$ denote the delay of a packet through queue, $\cq_3$, as
defined above. Then, \[ E[D_3] = O(n \log n).\]
\end{lemma}
\begin{proof}
An arrival occurs to $\cq_3$ when S and R meet with probability $0.5$.
Let $\{X_i\}$ be the sequence of inter-arrival times to this
queue. Then, $X_i$ are i.i.d. with $E[X_1] = 2 E[Z] = 2n$ and
$E[X_1^2] = \Theta(E[Z^2]) = \Theta(n^2\log n)$ from Lemma
\ref{lem:var}. The potential-departure process is an i.i.d. Bernoulli 
process with parameter $1/1.5n$. Let $\{Y_i\}$ be the sequence of
service times then $Y_i$ is a Geometric random variable with mean
$1.5n$. Hence $E[Y_1] = 1.5n$ and $E[Y_1^2] = \Theta(n^2)$. By
construction, the service process is independent of the arrival
process and hence $\cq_3$ is a GI/GI/1 FCFS queue. Then, by Kingman's
upper bound \cite{Wolff} on the expected delay for a GI/GI/1 -- FCFS
queue, the expected delay of $\cq_3$ is upper bounded as
\begin{eqnarray}
E[D_3] & = & O\left(\frac{E[X_1^2] + E[Y_1^2]}{E[X_1]}\right)~ = ~
O\left(\frac{n^2 \log n + n^2 }{n}\right) ~ = ~ O\left(n \log
n\right). \label{fe2}
\end{eqnarray}
\end{proof}

\begin{lemma}\label{lx2}
Let $D_4$ denote the delay of a packet through queue, $\cq_4$, as
defined above. Then, \[ E[D_4] = O(n \log n).\]
\end{lemma}
\begin{proof}
Consider the service process of $\cq_4$, which is $1$ at a potential
departure instant and $0$ otherwise. This is a stationary, ergodic
process since the inter-potential-departure times are i.i.d. with mean
$n$. The Bernoulli arrival process to $\cq_4$ is independent of the
service process with mean inter-arrival time $1.5n$. Since the arrival
and service processes form a jointly stationary and ergodic process
with mean service time strictly less than mean inter-arrival time, the
queue has a stationary, ergodic distribution with finite expectation
as shown by \cite{loynes}. Thus $\cq_4$ is stable.

Let $\tilde{Q}_t$ be the number of packets in the queue in time-slot
$t$ and let $Q_i$ be the number of packets in the queue at potential
departure instant $i$. Thus the process $\{Q_i\}$ is obtained by
sampling $\{\tilde{Q}_t\}$ at potential departure instants. Let
$A_{i+1}$ be the number of arrivals between potential departure
instants $i$ and $i+1$. Then the evolution of $Q_i$ is given by
\begin{equation}
Q_{i+1} = Q_i - {\mathbf 1}_{\{Q_i > 0\}} + A_{i+1}.
\label{e:Qevolution}
\end{equation}
Comparing the evolution of the process $\{Q_i\}$ with that of
$\{\tilde{Q}_t\}$ shows that $\{Q_i\}$ also has a stationary, ergodic
distribution. Let $Z$ be the inter-meeting time of any two nodes as
defined in the beginning of this section. Then since the arrival
process is Bernoulli and the inter-potential departure times are
i.i.d. with common distribution that of $Z$, it is clear the $\{A_i\}$
is a stationary process. Let $\tilde{Q}$, $Q$ and $A$ be random
variables with the common stationary marginals of $\{\tilde{Q}_t\}$,
$\{Q_i\}$ and $\{A_i\}$ respectively. Then taking expection in
(\ref{e:Qevolution}) under the stationary distribution, we obtain
\begin{equation}
P(Q>0) = E[A].
\label{e:PQnot0}
\end{equation}
The arrival proces is i.i.d. Bernoulli and hence conditioned on $Z$,
the distribution of $A$ is Binomial$\,(Z,2/3n)$. Since $E[Z]=n$ from
Lemma~\ref{lem:var}, we obtain
\begin{eqnarray}
E[A] &=& E[E[A|Z]] ~ = ~ E\left[ \frac{Z}{1.5n} \right] ~ = ~ 2/3.\label{e:Earrivals}
\end{eqnarray}
Squaring (\ref{e:Qevolution}), taking expecation, using the independence 
of $Q_i$ and $A_{i+1}$ and then rearranging  terms, we obtain
\begin{equation}
2(1-E[A])E[Q] = P(Q>0) + E[A^2] - 2E[A]P(Q>0).
\label{e:EQstep1}
\end{equation}
Using (\ref{e:PQnot0}) and (\ref{e:Earrivals}) in the above, we obtain
\begin{eqnarray}
E[Q] &=& \frac{E[A] + E[A^2] -2E[A]^2}{1(1-E[A])} ~=~ \frac{3}{2}\left(E[A^2] - \frac{2}{9}\right).
\label{e:EQ_EA2}
\end{eqnarray}
Recall that conditioned on $Z$ the distribution of $A$ is Binomial$\,(Z,2/3n)$ and hence
\begin{eqnarray}
E[A^2] &=& E[E[A^2]Z]] ~=~ \frac{2E[Z]}{3n} + \frac{4}{9n^2}\left(E[Z^2] -E[Z] \right)\nonumber\\
&=& \left(\frac{2}{3} - \frac{4}{9n} \right) + \frac{4}{9n^2}\Theta(n^2\log n) ~=~ \Theta(\log n), \label{e:EAsquared}
\end{eqnarray}
where we used Lemma~\ref{lem:var}. As a result it follows from
(\ref{e:EQ_EA2}) that
\begin{equation}
E[Q] = \Theta(\log n).
\label{e:EQoflogn}
\end{equation}
 
Next, we will bound $E[\tilde{Q}]$ using $E[Q]$.  To this end,
consider a time-slot $t$ and let the number of potential departures
before time-slot $t$ be $I(t)$. Thus time-slot $t$ is flanked by
potential departures $I(t)$ and $I(t)+1$. Then $\tilde{Q}_t
\leq Q_{I(t)} + A_{I(t)+1}$. Also using the fact that
$\{\tilde{Q}_t\}$ is ergodic, with probability $1$, we have
\begin{eqnarray}
E[\tilde{Q}]  &=& \lim_{T\tends\infty} \frac{1}{T} \sum_{k=1}^{T} \tilde{Q}_k \nonumber\\
&\leq&  \lim_{T\tends\infty} \frac{1}{T} \sum_{j=1}^{I(T)+1} \left(
Q_j Z_{j+1} + A_{j+1}Z_{j+1}\right)\nonumber \\
&=&  \lim_{T\tends\infty} \frac{I(T)+1}{T} \frac{1}{I(T)+1} \sum_{j=1}^{I(T)+1} 
\left( Q_j Z_{j+1} + A_{j+1}Z_{j+1} \right) \nonumber \\
&=& \frac{1}{E[Z]} \left( E[Q_1 Z_2] + E[A_1 Z_1]\right)\label{e:IbyT_renewal}\\
&=& \frac{1}{n} \left( E[Q]E[Z] + \frac{2}{3n}E[Z^2]\right) \label{e:QindZ}\\
&=& O(\log n).
\end{eqnarray}
We used the fact that $I(T)/T \tends 1/E[Z]$ by the elementary renewal
theorem \cite{Wolff} in (\ref{e:IbyT_renewal}) and the independence of
$Q_j$ and $Z_{j+1}$ in (\ref{e:QindZ}). Now using Little's formula,
since the arrival rate is $2/3n$, we conclude that
\[ E[D_4] = E[A]E[\tilde{Q}] = \frac{3n}{2}O(\log n) = O(n\log n).\]

\end{proof}

\section{Conclusion}
\label{sec:disc}

In this paper, we studied the maximal throughput scaling and the
corresponding delay scaling in a random mobile network with restricted
node mobility. In \cite{DGT}, it was shown that a particular mobility
restriction does not affect the throughput scaling.  In this paper, we
showed that it does not affect delay scaling either. In particular, we
show that delay scales as $D(n) = \Theta(n\log n)$ for a network of
$n$ nodes, which is the same as the delay scaling without any mobility
restriction. This was understood to be a consequence of the fact that
in spite of an apparent restriction, essentially the node mobility
remaining unchanged in the sense that (i) each node meets every other
node for $\Theta(1/n)$ fraction of the time with only $\Theta(1)$
other neighboring nodes, and (ii) the inter-meeting time of nodes
has mean of $\Theta(n)$ and variance of $O(n^2 \log n)$.

\end{document}